\documentclass[fleqn,10pt]{wlscirep}
\usepackage[utf8]{inputenc}
\usepackage[T1]{fontenc}

\usepackage{booktabs}
\usepackage{multirow}
\usepackage{longtable}
\usepackage{xr-hyper}
\usepackage[charter,cal]{mathdesign}

\makeatletter
\newcommand*{\addFileDependency}[1]{
  \typeout{(#1)}
  \@addtofilelist{#1}
  \IfFileExists{#1}{}{\typeout{No file #1.}}
}
\makeatother

\newcommand*{\myexternaldocument}[1]{%
    \externaldocument{#1}%
    \addFileDependency{#1.tex}%
    \addFileDependency{#1.aux}%
}
\myexternaldocument{SI}

\newcommand{\dl}{{\mathcal{L}}}

\title{
Human mobility is well described by closed-form
gravity-like models learned automatically
from data
}

\author[a]{Oriol Cabanas-Tirapu}
\author[a]{Lluís Danús}
\author[b,c,d]{Esteban Moro}
\author[a,*]{Marta Sales-Pardo}
\author[a,e,*]{Roger Guimerà}

\affil[a]{Department of Chemical Engineering, Universitat Rovira i Virgili, 43007 Tarragona, Catalonia}
\affil[b]{Institute for Data, Systems, and Society, Massachusetts Institute of Technology, Cambridge, MA 02139}
\affil[c]{Department of Mathematics and GISC, Universidad Carlos III de Madrid, 28911 Leganés, Spain}
\affil[d]{Network Science Institute, Northeastern University, Boston, MA 02115, United States}
\affil[e]{ICREA, 08007 Barcelona, Catalonia}

\affil[*]{Corresponding authors: Marta Sales-Pardo (E-mail: marta.sales@urv.cat); Roger Guimerà (E-mail: roger.guimera@urv.cat)}

\begin{abstract}
 Modeling of human mobility is critical to address questions in urban planning and transportation, as well as global challenges in sustainability, public health, and economic development. However, our understanding and ability to model mobility flows within and between urban areas are still incomplete. At one end of the modeling spectrum we have simple so-called gravity models, which are easy to interpret and provide modestly accurate predictions of mobility flows. At the other end, we have complex machine learning and deep learning models, with tens of features and thousands of parameters, which predict mobility more accurately than gravity models at the cost of not being interpretable and not providing insight on human behavior. Here, we show that simple machine-learned, closed-form models of mobility are able to predict mobility flows more accurately, overall, than either gravity or complex machine and deep learning models. At the same time, these models are simple and gravity-like, and can be interpreted in terms similar to standard gravity models. Furthermore, these models work for different datasets and at different scales, suggesting that they may capture the fundamental universal features of human mobility.
\end{abstract}
\begin{document}

\flushbottom
\maketitle

\section*{Introduction}
Accurate models of population mobility within and between municipalities are critical to address questions in urban planning and transportation engineering. Additionally, since municipalities are the main ground on which societies and cultures develop today, such mobility models are also instrumental in addressing global challenges in sustainability, public health, and economic development. Two main factors have driven recent interest in modeling human mobility patterns \cite{zipf46,erlander1990gravity,guimera05b,simini12,markus2021}. First, accurate models of human mobility could help identify transportation needs \cite{yuan2021survey}, allocate services and amenities (shopping, health, parks) more efficiently \cite{kingsley}, or even understand and eventually alleviate problems like segregation \cite{moro2021}, or epidemic spreading \cite{balcan2010modeling}. But, at the same time, models of human mobility can help identify the main behavioral components driving people to make large displacements to, for example, buy a new product, find a new house, or use physical activity spaces. Better behavioral models can help us implement more efficient policies to change people's behavior,  rather than urban environments, in favor of more sustainable attitudes.

\begin{figure*}[bt!]
\centering
\includegraphics[width=0.8\textwidth]{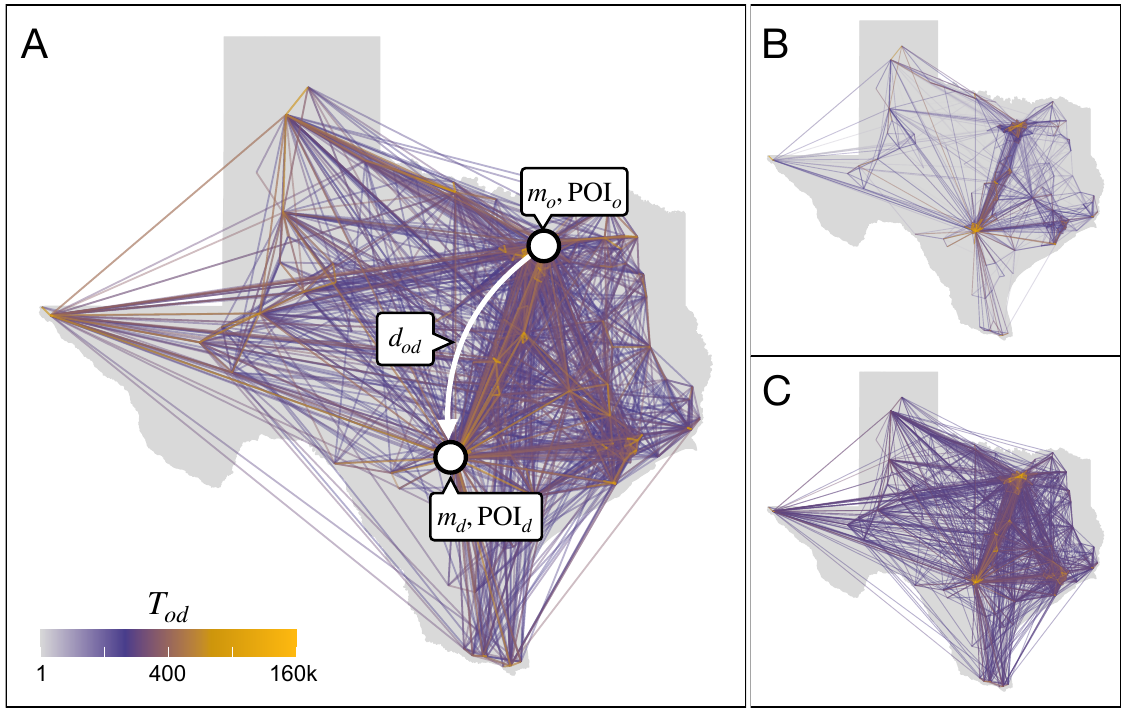}
\caption{
    {\bf Modeling approaches for mobility flows between test municipalities in Texas, US.}
    {\bf (A)} Real mobility flows between municipalities in the test set in Texas, US (Methods). For each flow, we consider origin $o$ and destination $d$ features, such as population $m_{o/d}$, aggregate statistics about points of interest (POI), and the distance between them.
    {\bf (B)} Flows predicted by the deep gravity model~\cite{simini21}, which uses a total of 39 features from origin and destination.
    {\bf (C)} Flows predicted by the closed-form, median predictive model identified by the Bayesian machine scientist (BMS; see text). This model only uses the population of origin and destination, as well as the distance between them (Fig.~\ref{fig:figure5}{\bf B}).
}
\label{figure1}
\end{figure*}

Despite these considerations, our understanding of the mobility flows within and between urban areas is still incomplete. One of the earliest and most fruitful attempts to model mobility flows between municipalities is the so-called gravity model \cite{zipf46}. This model assumes that mobility flows depend solely on the attractiveness or opportunities of the municipalities of origin and destination (for which population is typically used as a proxy) and the geographical distance between them, in a fashion that is mathematically similar to Newton's law of gravitation.
In its different incarnations and refined versions \cite{pappalardo2016, simini12,chen15}, the gravity model provides a simple phenomenological description of a very complex phenomenon. Because of this, while gravity models are not without their limitations, they are very often used in urban design, transportation, or even commercial applications. Recently, deep learning algorithms have been proposed, extending the ideas underlying gravity models; they incorporate many other features besides the populations of the origin and destination municipalities and their distance \cite{simini21}. Although those sophisticated machine learning tools are more accurate at predicting flows between urban areas, they lack the explanatory power, analytical tractability, adaptability to different contexts, and connection to human decision-making of simple gravity models.

Given the reasonable success of simple gravity models in explaining flows in urban areas, here we investigate the fundamental question of whether we really need non-interpretable models that are much more complex than the gravity law to delve deeper into the essence of urban mobility.

Unlike other behavioral models, gravity mobility models are phenomenological. 

Because of the lack of precise theoretical underpinnings, their predictive ability depends on the exact functional specification of the dependency of the mobility flows on the model features; that is, the mathematical dependency on origin and destination populations and distance. Here, we leverage recent developments in Bayesian symbolic regression to obtain closed-form, interpretable models\cite{rudin19} of mobility from data in a principled and automatic fashion \cite{guimera20,reichardt20,fajardo-fontiveros23}. 

In particular, we systematically compare the performance at predicting mobility flows of simple gravity models, complex machine learning and deep-learning methods, and closed-form, interpretable models obtained through Bayesian symbolic regression (Fig.~\ref{figure1}). We find that the Bayesian symbolic regression approach yields simple models that perform better, overall, than any of the other modeling approaches. Our approach is able to learn accurate models that, like gravity models, solely take into account the origin and destination populations and the geographical distance between them. Importantly, the learned models are gravity-like in their mathematical dependencies on populations and distance.

Furthermore, exploration of the relationship between the contribution of the populations of municipalities and their relative distance reveals common patterns in all the datasets, which suggests a close to universal relationship between mobility flows and these variables.

\section*{Results}

\subsection*{A Bayesian machine scientist learns closed-form mathematical models from mobility data}

We aim to determine whether it is possible to model mobility flows by means of closed-form mathematical models that are interpretable like gravity models, and as predictive as (non-interpretable) machine learning models such as the deep gravity model~\cite{simini21}. 
To automatically learn such closed-form models from data, we use the so-called Bayesian machine scientist (BMS)\cite{guimera20}. Given a dataset $D$, the BMS samples closed-form mathematical models from the posterior distribution $p(M|D)$, which gives the probability that a given model $M$ is the true generating model given the data (Methods). The BMS is guaranteed to asymptotically identify the true generating model, if one exists, and makes quasi-optimal predictions for unobserved data \cite{fajardo-fontiveros23}.

We consider as our main dataset $D$ the set of flows $T_{od}$ between origin $o$ and destination $d$ municipalities in six states in the USA (New York, Massachusetts, California, Florida, Washington, and Texas; see Data). The BMS is fed with $D$, and samples closed-form models from $p(M|D)$ using Markov chain Monte Carlo\cite{guimera20} (Methods). This sampling yields an ensemble of hundreds of different closed-form models for the flows $T_{od}$ such as, for example, 
\begin{equation}
    \log T_{od} = A \left( 1 + \frac{B\left( \left(m_d + C \right) \left( m_o + D \right)\right)^\beta}{d_{od}}\right)^\xi \quad {\rm or} \quad \log T_{od} = \log \left[ A \left( \frac{B\left( m_d m_o + C m_d + D\right)}{d_{od}^\alpha} + 1 \right)^\gamma\right]\;,
    \label{eq:mdl_model}
\end{equation}
where $m_{o/d}$ is the population of the origin/destination municipalities, $d_{od}$ is the distance between them, and $A$, $B$, $C$, $D$, $\beta$ and $\xi$ are model parameters. These models are able to make predictions of test flows (not seen by the BMS during training) that follow real values over several orders of magnitude (Figs.~\ref{figure1} and \ref{figure2}M-R). In what follows, we analyze in more depth this ensemble of models and its predictive abilities, vis a vis gravity models and machine learning models such as the deep gravity model.

\subsection*{Different models capture flows at different scales}

\begin{figure*}[bt!]
\centering
\includegraphics[width=0.95\textwidth]{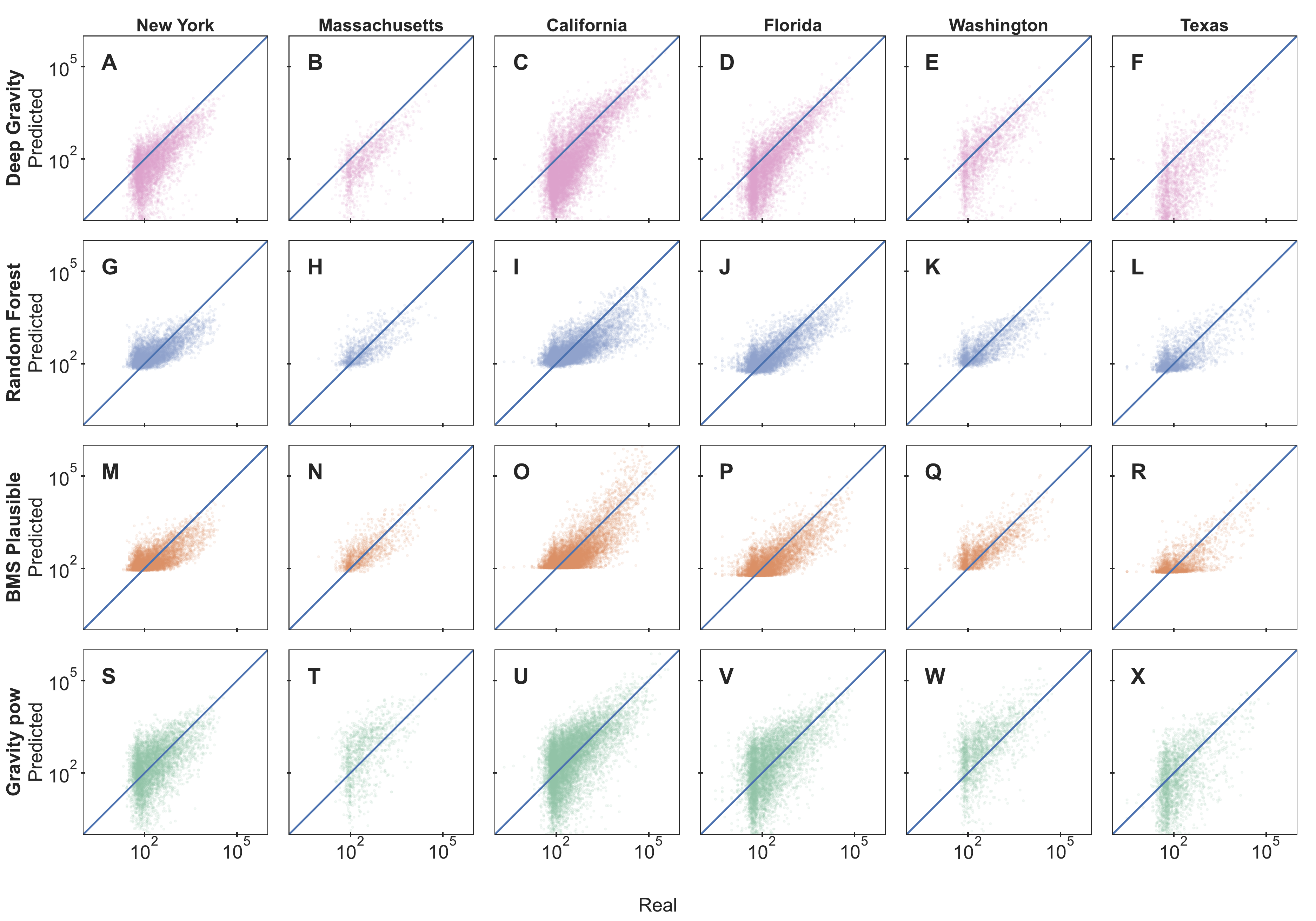}
\caption{\textbf{Model predictions of flows between municipalities.} Each panel shows, in logarithmic scale, the scatter plot of predicted flows between municipalities versus the corresponding real flows, for different states in the US (columns). Plots show results for test data for different models (rows): (\textbf{A-F}) the Deep Gravity model, (\textbf{G-L}) a Random Forest regressor, (\textbf{M-R}) the most plausible model sampled by the Bayesian machine scientist, and (\textbf{S-X}) a gravity model in its power law version. Supplementary Fig. S1 shows scatter plots for the full set of models we consider (see Methods for a complete description of the models and their parameters).}
\label{figure2}
\end{figure*}

In order to compare the ability of modeling approaches to describe mobility data, one needs a model selection criterion. In probabilistic terms, selecting the best model amounts to selecting the most plausible model, that is, the model that has the highest probability $p(M|D)$ of being the true generating model given the observed data; or, equivalently, the model with the shortest description length (Eqs.~\eqref{eq:boltzmann_dl} and \eqref{eq.dl}). However, this criterion is not always applicable in practice, because often it is not possible to compute the description length of a model, as it happens for deep learning and most other machine learning models.

Alternatively, one can measure performance at certain predictive tasks\cite{valles-catala18}, which is the approach typically taken in mobility modeling studies, and that we take here. 

Specifically, for each state for which we have data, we split municipalities in two sets. Flows between municipalities in the first set comprise the training set, and flows between municipalities in the second set comprise the test set. By building the training and test sets in this way, all the information about the municipalities in the test set, their characteristics, and the distances between them is completely new to the trained algorithm.

We compare the closed-form mobility models identified by the BMS to two alternative approaches. First, we consider gravity models, in which mobility flows are directly proportional to the product of masses (that is, populations) at the origin and destination, and inversely proportional to the distance between them. These approaches include traditional gravity models \cite{zipf46}, as well as the closely related radiation model \cite{simini12}. Second, we consider machine learning  approaches that, besides considering population and distance between municipalities, also consider additional characteristics of municipalities such as density of shops, entertainment venues, or educational facilities (Methods). Specifically we consider a random forest regression model \cite{ho95} and the deep gravity model \cite{simini21}.

From the ensemble of closed-form models sampled by the BMS, we analyze (Methods): (i) the most plausible (minimum description length) model found by the BMS; (ii) the median of the ensemble of models sampled by the BMS (which is the optimal predictor); and (iii) what we call the median predictive model, that is, the single model in the ensemble of sampled models whose predictions are closest to the ensemble median. 

In Fig.~\ref{figure2}, we show the predicted flows versus the real flows in the test set (see Supplementary Fig. S1 for results for additional models). Whereas all models are predictive, we find that different models are differently capable to describe mobility flows of different orders of magnitude. For instance, gravity-like models (Fig.~\ref{figure2} {\bf S-X} and Supplementary Figs.~S1 and S2 ) are typically good at capturing the behavior of large flows, but not of small flows. Indeed, for small flows (less than around 100 commuters) these approaches tend to underestimate flows, in some cases by several orders of magnitude, and even predict flows smaller than 1 person (Supplementary Fig. S2). This is also the case for the deep gravity model, which again under-predicts small flows (Figs.~\ref{figure1} and \ref{figure2} {\bf A-F}). By contrast, neither the random forest nor the BMS  suffer from this caveat, and both capture the whole range of flows more consistently and without large systematic deviations (Figs.~\ref{figure1} and \ref{figure2} {\bf G-R}).

\subsection*{Simple closed-form models are overall more accurate than gravity and non-interpretable machine learning models}

\begin{figure*}[bt!]
\centering
\includegraphics[width=0.95\textwidth]{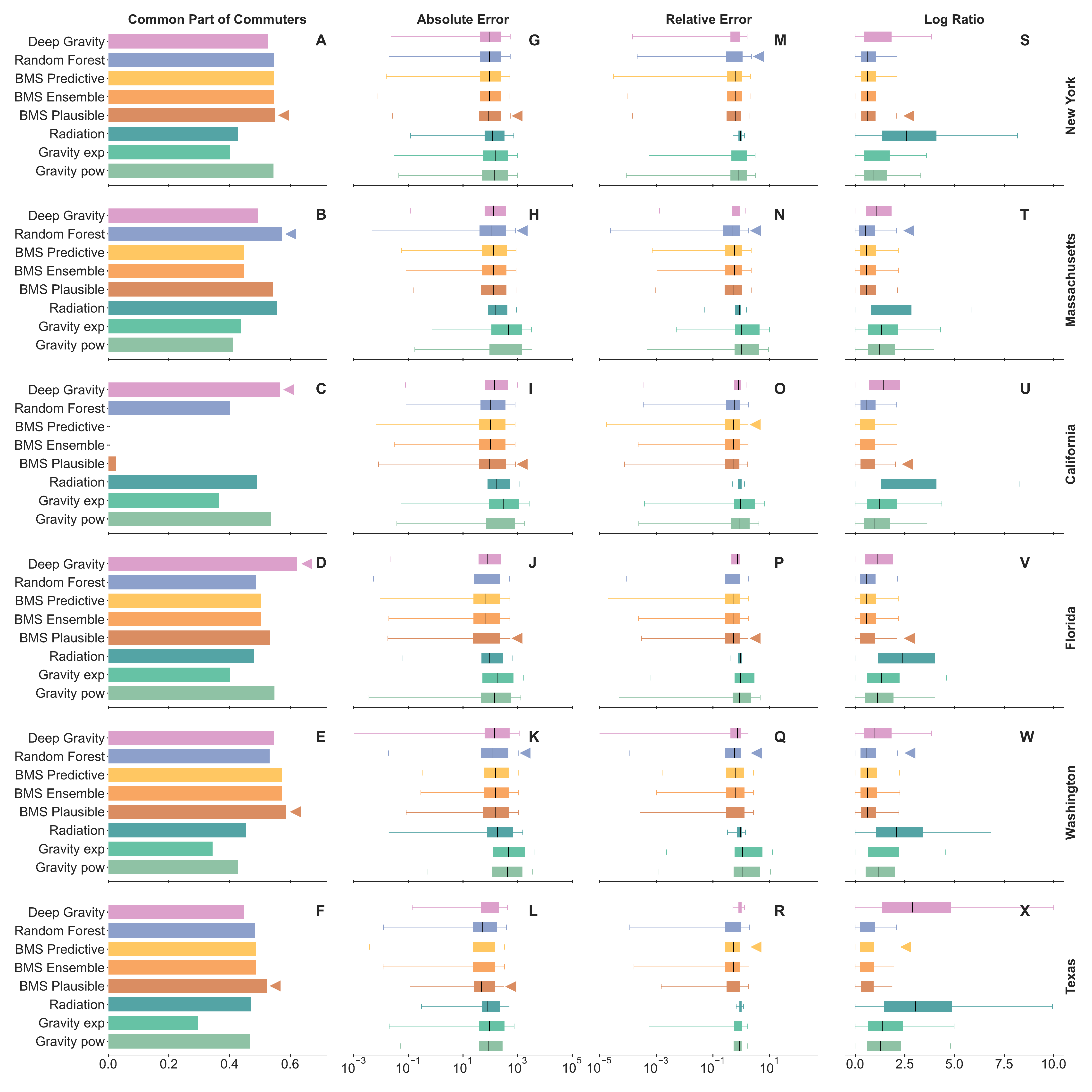}
\caption{\textbf{Model performance at predicting flows between municipalities.} For each one of the model predictions shown in Fig.~\ref{figure2} and Supplementary Fig.~S1 we assess model performance using four different metrics:\textbf{A-F}, Common part of commuters; {\bf G-L}  Absolute error; \textbf{M-R},  Absolute relative error; \textbf{S-X}, Absolute log-ratio. Each row corresponds to a different US state as indicated. The Common Part of Commuters is a global metric, thus we have a single value for each metric. For the other three metrics, we show the median, 50\% confidence interval (box) and 95\% confidence interval (whiskers).
Triangles ($\blacktriangleleft$) indicate the best performing model for each metric (largest CPC or lowest median).
See Methods and text for the definition and discussion of the different metrics. See Supplementary Table S3 for numerical values.}
\label{figure3}
\end{figure*}

Next, we quantify the performance of the models at the task of predicting unobserved flows. To that end, and considering the qualitative results in the previous section, we compute several complementary performance metrics. First, we consider the common part of commuters (CPC; see Methods), which is a usual choice in the mobility literature\cite{simini21}. The CPC measures the overlap between predicted and observed flows, and can take values from 0 to 1; the larger the CPC, the better the predictions. Despite its popularity, this metric favors models that predict the larger flows well, but overlooks errors in small flows (Fig.~\ref{figure3}A-F). Since mobility flows typically span several orders of magnitude (Fig.~2), models with the larger CPC are not necessarily the best models for the whole range of flows. 

To have metrics of performance that cover the whole range of flows, we consider, in addition and complementary to CPC, the absolute error, the absolute relative error, and the absolute log-ratio (Fig.~\ref{figure3}). For each of these metrics, and to avoid the disproportionate influence of singular large errors (especially for non-relative quantities such as the absolute error), we always show the whole distribution of error values (as a boxplot), and use the median value to compare models (Fig.~\ref{figure3}); the lower the median, the better the performance of the model.
Note that these metrics highlight different aspects of the prediction. Absolute errors are correlated with the magnitude of the flow we are trying to predict, so that errors are typically larger for larger flows. Because of this, and similar to the CPC, average absolute errors are very sensitive to the errors in predicting large flows but not to errors in small flows. 
For the same reason, median values of the absolute error typically reflect errors in performance for typical flow values, and do not reflect the ability of a model to predict values in the whole range of flows.   

The absolute relative error and the absolute log-ratio do take into account the effect of the magnitude of the flow, and therefore are more informative of the global behavior of a model when the range of flows spans several orders of magnitude (Fig.~\ref{figure3}M-X). An issue with the relative error is that while it penalizes over-prediction, it does not penalize under-prediction; in the extreme case in which the predicted flow equals zero and the real flow is larger than zero, the relative error is equal to one. As a result, distributions for relative errors in gravity-like models and the deep gravity model, in which small flows are under predicted, are centered around 1 (Fig.~\ref{figure3}S-X; Supplementary Fig.~S2). By contrast, the absolute log-ratio has the property that over- and under-prediction are equally penalized (in a logarithmic scale), that is,  predicting the real flow multiplied or divided by the same factor results in same absolute log-ratio. This metric therefore captures the ability of a model to predict flows in any range of values. 
 
\begin{figure*}[bt!]
\centering
\includegraphics[width=0.95\textwidth]{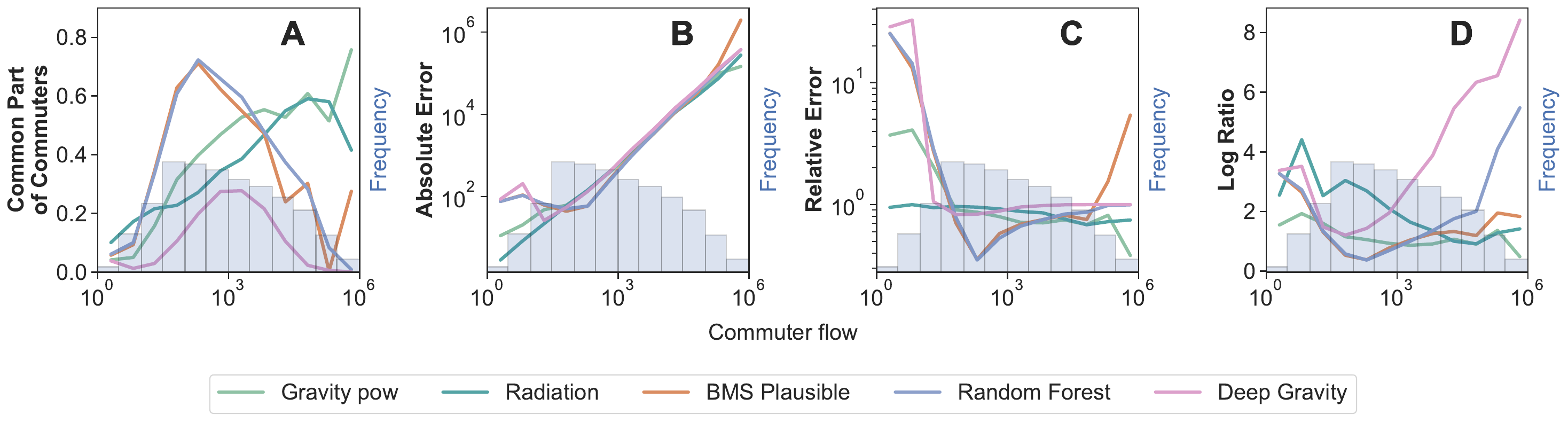}
\caption{\textbf{Performance for different flow ranges.} 
The observed flows between municipalities span six orders of magnitude and are distributed, over the six states we consider, as shown by the histogram. We measure the performance of the different models at predicting flows within each bin of flows: {\bf A}, CPC; {\bf B}, absolute error; {\bf C}, absolute relative error; {\bf D}, absolute log ratio.}
\label{figure4}
\end{figure*}

Using these metrics, we compare the different modeling approaches (Fig.~\ref{figure3}). The first conclusion from this comparison is that gravity models, including the radiation model, are never optimal; for all states and metrics we consider, there is always at least one other model that performs better. This is not surprising, since these models are simple and highly stylized, and they have already been shown to make less accurate predictions than deep gravity models\cite{simini21}.

Perhaps more surprisingly, we find that closed-form mathematical models obtained by the BMS perform, overall, at least as well or even slightly better than machine learning models, even when the latter are much more complex and can use many features other than population and distance. For the CPC, machine learning models are the best performing model for three out of six states (the random forest is best in two states, and deep gravity in one), whereas closed-form models identified as most plausible by the BMS are optimal in the three remaining states. In the case of California, the BMS overestimates a few extremely large flows (hence the CPC$\simeq 0$), but is able to describe all other flows well.

Similarly, for the relative error, random forest models are optimal in half of the States, whereas closed-form BMS models are optimal in the other half. When models are compared in terms of both absolute error and log-ratio, we find that closed-form BMS models are optimal in four of the States, whereas random forest models are optimal in the other two. Remarkably, deep gravity models are never optimal according to absolute error, relative error or log ratio.  

When considering how each model performs for flows in specific ranges (Fig.~\ref{figure4} and Supplementary Figure S3), we find that BMS models, similar to random forest models, are particularly good at modeling flows in the range that is the most common in the data.

Taking into account that both random forest and deep gravity models use many more features for their predictions (39 features in total, in contrast to the three features used by gravity models and the closed-form models identified by the BMS, namely, origin and destination population, and origin-destination distance), we conclude that the symbolic regression approach using the BMS yields more parsimonious models of human mobility flows between municipalities. Closed-form models obtained by the BMS also compare well to the alternatives in terms of the fairness of their predictions\cite{liu23} (Supplementary Fig. S4). In particular, we find that the random forest and the models identified by the BMS are, overall, the most consistent models across states in terms of the fairness of their predictions.

\begin{figure*}[bt!]
\centering
\includegraphics[width=0.95\textwidth]{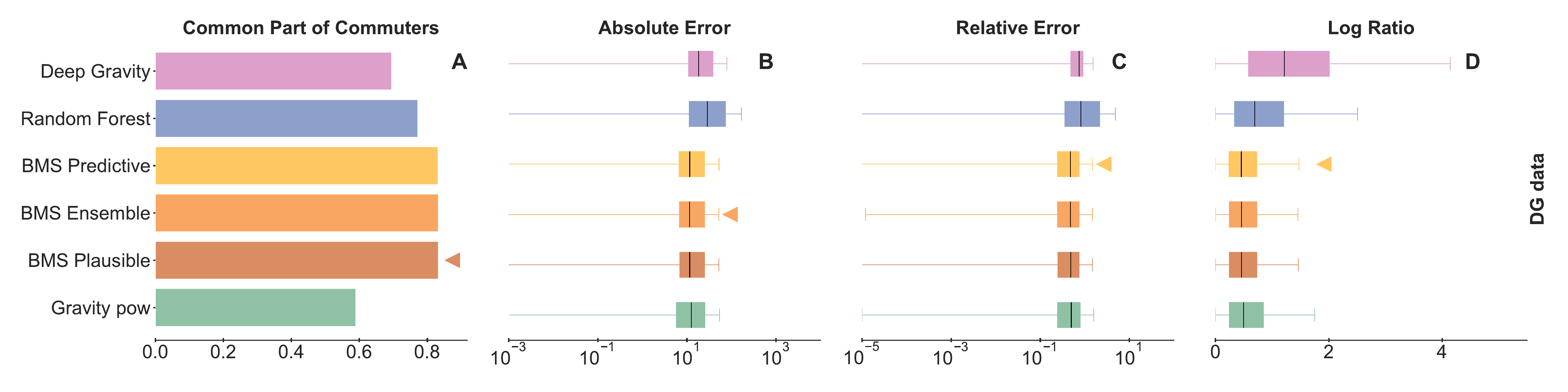}
\caption{\textbf{Model performance at predicting flows at small distances.}
We consider the data used by Simini et al.\cite{simini21} about mobility flows between census tracts within small geographical (25 km$^2$) regions in NY state. We evaluate predictions over test data using the same metrics as in Fig.~\ref{figure3}. Different columns correspond to different metrics and  triangles ($\blacktriangleleft$) indicate the best performing model according to each metric. See Methods and text, for a complete description of the metrics, the models, and the data.}

\label{metricsDG}
\end{figure*}

\subsection*{Closed-form models also describe flows at shorter scales}

So far, we have analyzed within-State flows between municipalities, at any range of distances and of any size. However, it may be that the geographical, economic and demographic characteristics of smaller areas become relevant when modeling flows at shorter distances (for example, within neighborhoods or adjacent towns in large metropolitan areas). To elucidate to what extent different modeling approaches can accommodate to such short-distance flows, we adopt the framework used in Ref.~\cite{simini21}: we divide the state of New York in small tiles of 25 km$^2$, and consider the flows between census tracts within each tile.\cite{simini21} We use 50\% of the tiles for training the different models, and then test on the flows between the remaining 50\% of the tiles. For this experiment we find that, regardless of the metric used, closed-form models identified by the BMS are always more accurate than machine learning and gravity models (Fig.~\ref{metricsDG}). Our results thus indicate that simple closed-form models that just consider populations of municipalities/census tracts and the distances between them provide better descriptions of mobility flows than complex models that take many more features into account, and have many more parameters, also for flows at short distances.

\subsection*{The Bayesian machine scientist finds gravity-like models to describe mobility flows}

Our analysis indicates that the BMS is able to find closed-form mathematical models that solely consider the populations of the origin and destination and the geographic distance between them; and that these models provide predictions of mobility flows that are even more accurate than complex models such as the deep gravity model. Here, we investigate whether, besides being predictive, these closed-form models are also interpretable and insightful.

We start by noting, once more, that the BMS samples hundreds of models and that, overall, they all perform well, which shows that there are many different models that can describe the data. Such a set of models is sometimes called a Rashomon set\cite{rudin19}. Then, the relevant question is whether these models share any common defining properties that could explain why they describe mobility flows accurately.

\begin{figure*}[bt!]
\centering
\includegraphics[width=0.95\textwidth]{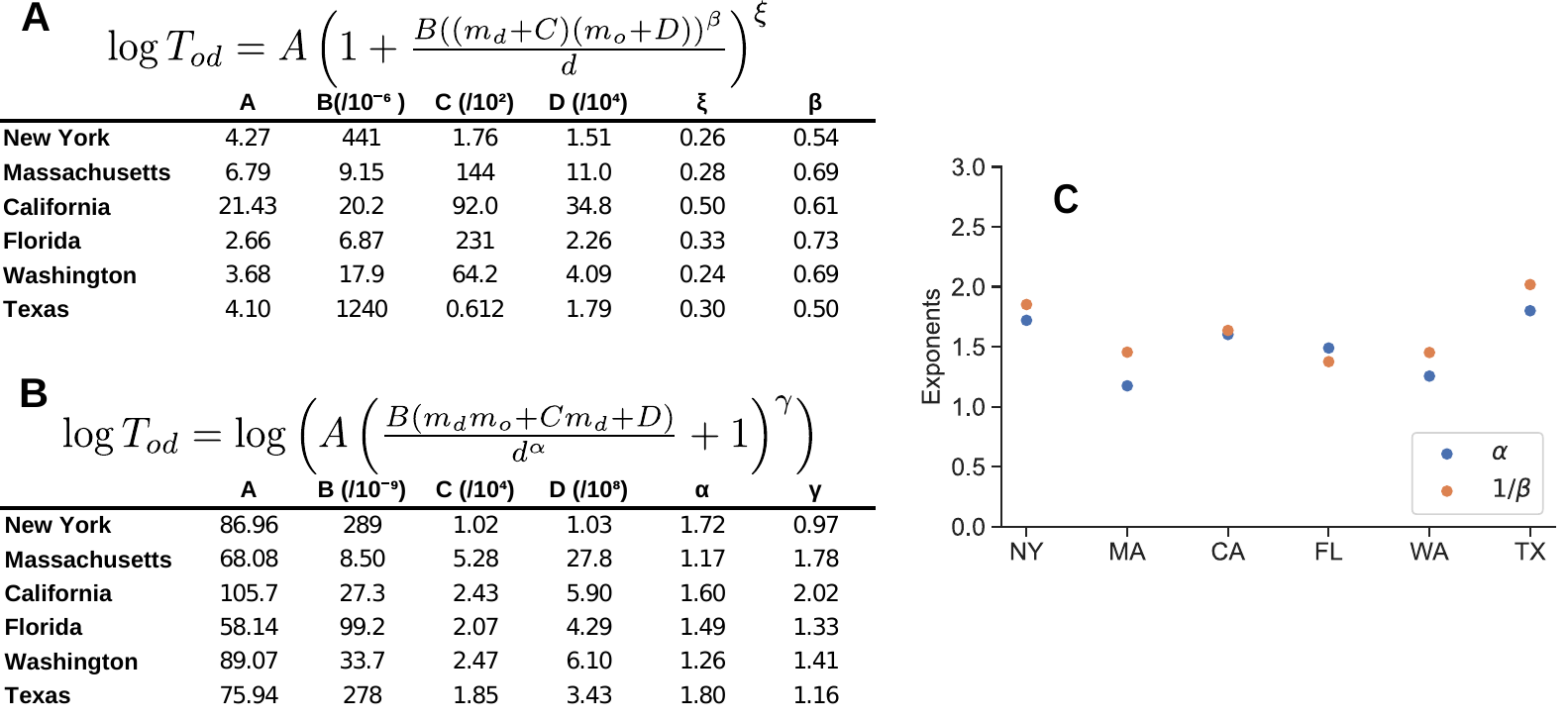} 
\caption[B]{\textbf{Closed-form models for mobility flows.} We ran the Bayesian machine scientist (BMS) with a training set of 1000 points and three features: origin and destination populations and distance between them. We used 5 independent Markov chains of 12,000 Monte Carlo steps each. (\textbf{A}) Minimum description length model (Methods) for the logarithm of the data, where $d$ is the inter-municipality distance, $m_o$ is the origin population, and $m_d$ is the destination population. In the table we show the fitting parameters for each state of the training data. (\textbf{B}) Median predictive model (Methods) for the logarithm of the data. As before, $d$ is the inter-municipality distance, $m_o$ is the origin population, and $m_d$ is the destination population. In the table we show the fitting parameters for each state of the training data. (\textbf{C}) Ratio between the distance exponent and the population exponent. Blue points are obtained from the most plausible model and orange points from the median predictive model.}
\label{fig:figure5}
\end{figure*}

To elucidate this question, we analyze two particularly relevant closed-form models identified by the BMS (Fig.~\ref{fig:figure5}): (i) the most plausible model, that is, the model that has the highest probability $p(M|D)$ given the data (or, equivalently, the shortest description length $\dl(M,D)$) among all those sampled by the BMS (Methods); (ii) the median predictive model, that is, the closed-form model whose predictions for unobserved data are closest to the median prediction of the whole ensemble of sampled models (Methods). Formally, there are marked mathematical differences between both models. In particular, the most plausible model is an exponential model for the flows, while the median predictive modesl is a power-law model for the flows. However, the two models have relevant properties in common. First, both models are gravity-like models, that is, they depend on a product of the origin and destination populations (shifted by a certain amount), and inversely on a function of the geographic distance between origin and destination. This is remarkable because the BMS has not received any input about the particular shape that models should take, which suggests that the regularities in the data are well-described by this general class of models and justifies the historical use of gravity models.

Second, we find that origin and destination do not necessarily play a symmetric role, which allows the model to accommodate non-symmetric flows in contrast to typical gravity models which do not allow for this possibility. Indeed, an inspection of the parameters shows that in some states such as  New York or Texas, flows are much more symmetric than in others such as Florida and Massachusetts. 

Finally, we also find that the relative contribution of the mass product with respect to the geographical distance is consistent across models. In both models, we find a mathematically equivalent dependency on the ratio $(m_d m_o)/d^e $, where $e=1/\beta$ in the most plausible model (Fig.~\ref{fig:figure5}{\bf A}), and $e=\alpha$ in the median predictive model  (Fig.~\ref{fig:figure5}{\bf B}). We find that, for a given state, $\alpha\approx 1/\beta$ suggesting that this relationship is to a large extent model-independent (Fig.~\ref{fig:figure5}{\bf C}). Furthermore, we find that the state-to-state variability is small since all exponents fall within the range $[1,2] $ (most within the range $[1.5,2]$), which suggests that reasonable models for mobility flows are gravity-like models with specific constraints in the relationship between the contributions of the mass product and the geographical distance.

\section*{Discussion}

Understanding human mobility is critical to address questions in urban planning and transportation, as well as global challenges in sustainability,
public health, and economic development. Traditionally, mobility flows have been modeled using simple gravity models, which are conceptually simple and easy to interpret, but have limited predictive power. Recently, deep learning models have been proposed as an alternative; whereas these models are consistently and significantly more predictive than gravity models, they are not interpretable and provide little insight into human behavior. Here, we have shown that automated equation discovery approaches lead to parsimonious models that combine the most desirable aspects of both approaches---the simplicity of gravity models and a predictive power that is even better, overall, than deep learning models.
Remarkably, the models we identified are gravity-like in that they are increasing functions of a certain product of populations of origin and destination, and decreasing functions of the distance between them. While the ratio between the population and the distance terms is, in principle, model and data-dependent, in the datasets we explore the ratio is roughly constant and dataset-independent. 

Individual mobility depends critically on the urban environment, personal preferences, commuting patterns, and accessibility to transportation and amenities. Thus, modeling mobility at an individual or small spatial scale might require more complicated models that account for routes, the purpose of the trip, points of interest, or even the demographic traits of individuals \cite{liu23,simini21,deepmove}. However, our results show that by aggregating mobility at a larger spatial scale, the movements of millions of people can be described by simple fully explainable models that do not depend on the microscopic characteristics of the origin, destination, or route taken. This is because the randomness and variability inherent in individual behaviors tend to cancel out when looked at collectively, revealing underlying trends and movements that are driven by the shared needs of large populations, and the structure of the built environment. Our results show, therefore, that the aggregated flows in human mobility can be seen as an emergent and universal property of the complex system of individual movements. More broadly, our results showcase the potential of using machine scientists\cite{dzeroski07,evans10,guimera20} to automate the process of finding similar phenomenological closed-form models from data; and to use these models to gain insight into the relevant variables and mechanisms to describe other complex phenomena.

\section*{Methods}


\subsection*{Mobility data between municipalities in the US}
We collected weekly flows between United States census tracts \cite{kang20} for the first week of January 2019 (2019/01/07 to 2019/01/13) and the first week of March 2019 (2019/03/04 to 2019/03/9). The data consist of anonymous mobile data trajectories, regardless of the transportation method. The data set contains the geographical identifier (GEO ID) for both origin and destination census tracts as well as their corresponding geographical coordinates, the estimated number of visitors detected by \href{https://www.safegraph.com/}{SafeGraph}, and the estimated population flows inferred from the number of visitors. The datasets are available at \href{https://github.com/GeoDS/COVID19USFlows-WeeklyFlows-Ct2019}{GeoDS}.

\paragraph{Data processing}
In order to obtain the flows between municipalities from the data using census tract data, we first match municipalities (cities, towns and villages) with their corresponding census tracts. Then, the total flow between two municipalities A and B is calculated as the aggregate of flows between the set of census tracts  municipality A is comprised of and the set municipality B is comprised of. 

Specifically, we consider mobility data sets within six states in the US: New York, Massachusetts, California, Florida, Washington, and Texas. For each state, our data consist of the origin-destination municipality names, the flow, the distance, and the origin-destination populations and POI categories. We only consider municipalities with a non-zero population and pairs of municipalities with non-zero flow; we do not consider flows within the same municipality (See Supplementary Table~\ref{t.summary} for details).

\paragraph{Information about municipalities and census tracts} We obtained shapefiles containing the geographical coordinates of the polygons delimiting census tracts and municipalities  from \href{https://www.census.gov/cgi-bin/geo/shapefiles/index.php?year=2021}{United States Census Bureau}. We also retrieved population data of each municipality using the GEO ID \href{https://docs.datacommons.org/api/}{Data Commons}.

Finally, using  a local copy of \href{https://www.openstreetmap.org}{Open Street Map (OSM)} with \href{http://overpass-api.de/}{Overpass API}, we retrieved information about Points of Interest (POI) in each census tract.  We selected 18 categories of OSM elements that represent geographical, demographic and socio-economic features of the different municipalities.

\paragraph{Construction of train and test datasets}
To generate the train and test datasets for each state, we divide municipalities in two approximately equal sets at random. The training fold comprises flows between municipalities in the first set; the test fold comprises flows between municipalities in the second set. By doing so, we ensure that we test the learned models on data never observed before (not only in terms of flows, but in terms of municipality features as well). Tables \ref{Train_datasets} and \ref{Test_datasets} show the characteristics of train and test datasets.
To speed up the training process, we train the models with the same random sample of 1000 points of the training fold except for the Deep Gravity model for which we have to use a lager number of data points for training. 
\begin{table}[h!]
\begin{center}
    \begin{tabular}{lrr|rr|rr|rr|}
                State & Entries & Municipalities &  \multicolumn{2}{|c|}{Flow} & \multicolumn{2}{|c|}{Distance (Km)} & \multicolumn{2}{|c|}{Population} \\
                 & & & Min & Max & Min & Max & Min & Max \\ \midrule
                New York & 1000 & 217 & 26 & 122991 & 1.83 & 489.80 & 185 & $8.80\cdot10^6$\\
                Massachusetts & 1000 & 75 & 19 & 66946 & 2.34 & 205.51 & 1029 & 675647 \\
                California & 1000 & 260 & 14 & 91267 & 2.12 & 1078.21 & 237 & $1.013\cdot10^6$\\
                Florida & 1000 & 223 & 7 & 78316 & 2.54 & 607.50 & 251 & 949611\\
                Washington & 1000 & 107 & 20 & 71796 & 2.17 & 467.16 & 20 & 228989\\
                Texas & 1000 & 177 & 21 & 192660 & 1.84 & 975.16 & 173 & $2.30\cdot10^6$\\
            
    \end{tabular}
    \caption{\textbf{Train dataset}. Number of points and municipalities in the train set for each state, obtained form a random sample of 1000 points of the original train fold. We also detail the lowest and largest flow, distance and municipality population.}
    \label{Train_datasets}
\end{center}
\end{table}

\begin{table}[h!]
\begin{centering}
    \begin{tabular}{lrr|rr|rr|rr|}
                State & Entries & Municipalities &  \multicolumn{2}{|c|}{Flow} & \multicolumn{2}{|c|}{Distance (Km)} & \multicolumn{2}{|c|}{Population} \\
                 & & & Min & Max & Min & Max & Min & Max \\ \midrule
                New York & 5952 & 249 & 20 & 35063 & 0.77 & 531.39 & 361 & 211569\\
                Massachusetts & 1180 & 75 & 9 & 55499 & 2.16 & 267.24 & 1517 & 206518 \\
                California & 11727 & 319 & 2 & 416696 & 1.09 & 1084.34 & 129 & $3.9\cdot10^6$\\
                Florida & 7092 & 245 & 3 & 79624 & 1.14 & 875.82 & 78 & 442241\\
                Washington & 2083 & 109 & 12 & 54245 & 1.48 & 431.53 & 487 & 737015\\
                Texas & 2782 & 192 & 3 & 112865 & 2.46 & 1196.33 & 106 & $1.43\cdot10^6$\\
            
    \end{tabular}
    \caption{\textbf{Test dataset}. Number of points and municipalities in the test set for each state. We also detail the lowest and largest flow, distance and municipality population. }
    \label{Test_datasets}
\end{centering}
\end{table}

\paragraph{Economic classification of municipalities}
We retrieve the median income per capita of each municipality and state as of 2020 from \href{https://docs.datacommons.org/api/}{Data Commons}. Then we classify each municipality with the label  \emph{rich} if the median income of the municipality is above the median income of the state, or \emph{poor} if the median income per capita of the municipality is below the median income of the state.

\subsection*{Mobility data at small scales from Simini {\it et al.} \cite{simini21}.} The code available for the \href{https://github.com/scikit-mobility/DeepGravity}{Deep Gravity model} provides data at the level of census tracts for the New York State area. Mobility flows are obtained from the same dataset. In order to obtain the predictions of the Deep Gravity model for each individual trip and also the same dataset, we run the program and save, for each trip, the corresponding tile, the real value, the prediction and the variables. We store the train and test sets for the comparison with other methods.

\subsection*{Bayesian Machine Scientist} 
The Bayesian machine scientist (BMS) is a Bayesian approach to symbolic regression that estimates the plausibility of a closed-form mathematical model $M$ given the observed data $D$ as the  posterior probability $p(M|D)$. Without loss of generality, this posterior can be written \cite{guimera20}
\begin{equation}
    p(M|D) = \frac{\exp \left[-\dl(M, D)\right]}{Z} \,,
    \label{eq:boltzmann_dl}
\end{equation}
where $\dl(M, D)$ is the description length\cite{grunwald07} of the model (and the data), and $Z=\sum_{M'}\exp \left[-\dl(M', D)\right]=p(D)$ is the evidence. Within standard approximations\cite{schwarz78,guimera20}, the description length can be computed as
\begin{equation}
    \dl(M, D) = \frac{B(M, D)}{2} - \log p(M) \,,
    \label{eq.dl}
\end{equation}
where $B(M, D)$ is the Bayesian information criterion\cite{schwarz78}, which is straightforward to calculate from the data, and $p(M)$ is a suitable prior distribution over models\cite{guimera20}.

In this work, our goal is to simultaneously model flows from six different states in the US (that is, six different datasets). To this end, we use a multi-dataset approach\cite{reichardt20} which consists in finding a unique closed-form model for multiple datasets.
For each dataset, we allow model parameters to take different values\cite{reichardt20}. For a single dataset $D=\{(y_i,{\bf x}_i)\}$, where $\{y_i\}$ is the set of observations and $\{{\bold x}_i\}$ is the set of feature values associated to each observation,
the description length of a closed-form model $M$ and the data is  given by Eq.~\eqref{eq.dl}.
In the case in which our data comprises $K$  independent datasets  $D=\{D_k,~ k=1\dots,K\}=\left \{\{(y^1_i,{\bold x}^1_i)\},\dots, \{(y^K_i,{\bold x}^K_i)\}\right\}$ that we want to model using a single model $M$, the description length is\cite{reichardt20}
\begin{equation}
  \dl(M,D) = \frac{1}{2}\sum_k B(M, D_k)  - \log p(M)~.  
  \label{eq.dl_k}
\end{equation}
The BMS represents closed-form models as labeled trees and uses Markov chain Monte Carlo (MCMC) to explore the space of closed-form mathematical models by sampling from the posterior distribution $p(M|D) \propto \exp (-\dl)$. In this work, we consider models for flows with three independent variables (populations at origin and destination and geographical distance) and up to six parameters.  

The ensemble of sampled models allows to make predictions on the test set using three different approaches:
\begin{enumerate}
    \item The most plausible model. This is the model with the shortest description length that the BMS is able to find. 
    \item The ensemble of models. Ensemble predictions are the optimal predictions since they correspond to fully integrating over model space.  We estimate this integral by averaging over the predictions made by each one for the models we sample. Specifically, we perform five independent realizations of 12,000 MCMC steps. We then collect a model every 100 steps within the last 2,000 steps of the Markov chain to obtain an ensemble of 100 models.
    \item The median predictive model. This is the model within the ensemble whose predictions are closest to the predictions of the ensemble as a whole.
\end{enumerate}

\subsection*{Benchmark models}

\paragraph{Gravity model.} We consider the gravity model in its simplest form \cite{zipf46} in which the observed flow $T_{ij}$ between municipalities $(i,j)$ is a function of the populations $m_i$ and $m_j$ of the municipalities and the distance $d_{ij}$ between them 
\begin{equation}
    T_{i j}= C\frac{m_i \,m_j}{f(d_{ij})} \,.
\end{equation}
Here $C$ is a scaling parameter and $f(d)$ is a function of the distance.
Specifically, we consider two possible choices for $f(d)$: i) a power-law $f_{\rm pow}(d) 
=d^{\alpha}$; and ii) an exponential-law $f_{\rm exp}(d)=\exp(\alpha d)$. In both cases, the parameter $\alpha$ is obtained by fitting the model to the data in the training set. Because flows span several orders of magnitude, we find that training the model on the logarithm of the flows, rather than the flows themselves, leads to more predictive models. Therefore, all results reported here for gravity models correspond to this approach.

\paragraph{Radiation model.} We consider the original formulation of the model \cite{simini12}, in which  flow $T_{ij}$ is modeled as the  the total outflow of an origin municipality $T_i$ times the probability of going from $i$ to $j$. This probability depends on the populations of the origin ($m_i$) and destination($m_j$), as well as the populations of the municipalities within a radius $d_{ij}$ from the municipality at the origin:

\begin{equation}
    T_{i j}= T_i\, p_{i\to j}= T_i \frac{m_i \,m_j}{(m_i + s_{i j})\,(m_i + m_j + s_{i j})},
\end{equation}
where  $s_{i j}=\sum_{k\neq i,j} m_k (\forall k \; d_{i k} < d_{i j})$.

Recent works introduce modifications to this model for finite-size systems and in order to avoid border effects \cite{masucci2013,yang14,simini12,lenormand2012,lenormand16}(See Supplementary Fig. S5). However, we find that these models do not outperform the original formulation . 

\paragraph{Random Forest. } We implement a Random Forest regressor \cite{scikit-learn} with 1,000 estimators. As input data we use a total of 39 features for each origin-destination pair which include: distance, population at origin and destination, and 36 geographical and socio-economic features of the origin and destination areas (see Data and Supplementary Table \ref{t.poi}).

\paragraph{Deep Gravity.} Taking as a baseline the algorithm developed in \cite{simini21}, we modified the model to predict flows between municipalities rather than between small geographic regions resulting from tessellation. The major difference with the original model is that municipalities are now the smallest geographic unit, allowing us to compare with the models evaluated in this study. In all other aspects (features used, pre-processing of data, an model training) the model remains the same. The modified version of the code can be consulted and downloaded here \href{https://github.com/ocabanas/Symbolic\_mobility\_BMS/DeepGravity}{https://github.com/ocabanas/Symbolic\_mobility\_BMS/DeepGravity}.

\subsection*{Metrics}

\paragraph{Common Part of Commuters (CPC).} It is a widely used metric to analyze the performance of mobility models that is defined as
\begin{equation}
    \mathsf{CPC}=\frac{2\sum_{ij}{\rm min}(T_{ij},T_{ij}^*)}{\sum_{ij}T_{ij}+T_{ij}^*} \; ,
\label{CPC}
\end{equation}
where $T_{ij}$ is the predicted value of the flow from $i$ to $j$ and $T^*_{ij}$ is the observed flow. The maximum value of the CPC is $1$ if there is a complete agreement between the real data and predictions and it decreases to $0$ if all predictions for any flow are equal to zero. Note that the CPC is biased toward models that make accurate predictions for large flows, since smaller flows have marginal contributions to the sums. This is especially critical in mobility data where flows can span several orders of magnitude (see Tables \ref{Train_datasets}, \ref{Test_datasets}).

\paragraph{ Absolute error.} It measures the distance between the real and predicted data
\begin{equation}
    E_{ij}=|T_{ij}-T_{ij}^*|~~.
\end{equation}
Note that absolute error scales with the size of the flows, so that the average absolute error is biased towards the errors of average flows. For this reason we represent the whole distribution.

\paragraph {Relative error.} It measures the difference between real and predicted flows relative to the observed value of the flow:
\begin{equation}
    \epsilon_{ij}=| \frac{T_{ij}-T_{ij}^*}{T_{ij}^*}| ~.
\end{equation}
Note that, while over-predicting flows is penalized by the relative error, under-predicting flows is not, since predicting a zero value for a non-zero flow  results in $\epsilon_{ij}=1$. Because this can bias average relative errors, we plot the whole distribution.

\paragraph{Absolute log-ratio.} It measures the difference in the logarithms of predicted and real flows:
\begin{equation}
    LR_{ij}=|\log \frac{T_{ij}}{T_{ij}^*}|~.
\end{equation}
Note that for a perfect prediction this metric is equal to zero. Importantly this metric penalizes both over- and under-predictions equally. For instance, a prediction of a flow twice as large $T_{ij}=2 T^*_{ij}$ has $LR_{ij}=\log~{2}$, and a prediction  $T_{ij}=T^*_{ij}/2$ has $LR_{ij}=\log~{2}$ as well.

\paragraph{Proportional Demographic Parity (PDP).} The goal of this metric is to quantify the fairness of a model when predicting flows between different demographic or socio-economic groups $\{g\in \mathcal{G}\}$. To do so, it quantifies to what extent the errors of the predictions for flows across pairs of groups $\{ f_{ij}\equiv (g_i,g_j) \in \mathcal {G}^2 \equiv \mathcal{G}\times \mathcal{G}\}$ are equally distributed for all pairs of (different) groups. 
Consider that $\bar{l}$ is the median error of all flows, and  $\tau$ is a percentile window around the median ($0\leq \tau \leq 100$). For a pair of flow groups $(f_1,f_2)$,  $\mathsf{PDP}$ estimates the difference between their error distributions as 
\begin{equation}
\mathsf{ PDP}_{f_1,f_2}= \left\vert P (\overline{l}-\frac{\tau}{2} \leq l \leq \overline{l}+\frac{\tau}{2} |  f_1) - P (\overline{l}-\frac{\tau}{2} \leq l \leq \overline{l}+\frac{\tau}{2} |  f_2) \right\vert
\label{eq.pdp_pair}    
\end{equation}
where $P(\cdot |f_i)$ is the probability that a prediction of a flow in flow group $f_i$ has an error $l$ such that $\overline{l}-\frac{\tau}{2} \leq l \leq \overline{l}+\frac{\tau}{2} $. 

To get an overall estimate, then $\mathsf{PDP}$ uses a weighted average\cite{liu23}
\begin{equation}
   \mathsf{PDP}= \sum_{f,h\in \mathcal{G}^2 ,f\neq h} w_{f,h} \, \mathsf{PDP}_{f,h}~~~, 
\end{equation}
where the weight $w_{f,h}=\sum_{k\in \mathcal{G}^2}N_{k}/\left(N_{f}+N_{h}\right)$ enhances the relative contribution of small groups of flows. Note that our approach to measure $\mathsf{PDP}$ is a generalization of that used in Ref.\cite{liu23}, where, instead of percentile windows, the authors consider $\tau$ to be a standard deviation around the mean. However, because error distributions are not Gaussian in general (see explanation for the different error metrics), we use a more general definition applicable to any distribution.  

In our analysis, we consider two groups of municipalities: above (rich) and below (poor) the median income per capita (see Data). Therefore we have four different  flow groups $\mathcal{G}^2=\{{\rm poor \to poor},\, {\rm rich \to poor},\,{\rm poor \to rich},\, {\rm rich \to rich}\}$.

\bibliography{mobility}

\section*{Acknowledgements}
We thank L. Pappalardo and M. Luca for help with the Deep Gravity algorithm, and M. Luca for comments and suggestions on the manuscript. This research was supported by projects PID2019-106811GB-C32 (E.M), PID2019-106811GB-C31 and PID2022-142600NB-I00 (M.SP. and R.G.), and FPI grant PRE2020-095552 (O.CT.) from MCIN/AEI/10.13039/501100011033, and by the Government of Catalonia (2021SGR-633) (M.SP. and R.G.). E.M. acknowledges support from the National Science Foundation under Grant No.~2218748.

\section*{Author contributions statement}
O.C. collected data.
O.C. and L.D. wrote code and performed experiments.
All authors designed research, analyzed results, discussed results, and wrote the paper.

\subsection*{Data and code availability}
All data are available as described in the Methods section.
Data used for the evaluation of the Deep Gravity model are available at \href{https://github.com/ocabanas/Symbolic_mobility_BMS}{Github}.
The code for the BMS is available from \href{https://bitbucket.org/rguimera/machine-scientist/}{https://bitbucket.org/rguimera/machine-scientist}.

\section*{Competing interests}
The authors declare no conflict of interest.

\end{document}


\flushbottom
\maketitle

\tableofcontents

\newpage


\section{Supplementary Tables}
\label{sectionSX:data}
\subsection{Data Summary}

\begin{table}[!ht]
\begin{centering}
    \begin{tabular}{l|r|r|rr|rr|rr}
                State & Entries & Municipalities &  \multicolumn{2}{|c|}{Flow} & \multicolumn{2}{|c|}{Distance (Km)} & \multicolumn{2}{|c}{Population} \\
                 & & & Min & Max & Min & Max & Min & Max \\ \midrule
                New York & 22492 & 498 & 6 & 267527 & 0.77 & 547.74 & 185 & 8804190\\
                Massachusetts & 5024 & 150 & 2 & 66946 & 2.16 & 269.79 & 1029 & 675647 \\
                California & 48532 & 644 & 2 & 508755 & 0.87 & 1183.85 & 129 & 3898747\\
                Florida & 26901 & 490 & 3 & 113048 & 0.73 & 909.38 & 78 & 949611\\
                Washington & 7644 & 219 & 12 & 77671 & 1.23 & 480.54 & 20 & 737015\\
                Texas & 13487 & 386 & 1 & 242726 & 0.84 & 1196.33 & 106 & 2304580\\
            
    \end{tabular}
    \caption{\textbf{Data Summary}. Number of nonzero flow entries and municipalities considered for each state.  We also detail the lowest and largest flow, distance and municipality population for each state.}
    \label{t.summary}
\end{centering}
\end{table}

\subsection{Point of interest (POI) categories}

To obtain the POIs of the selected categories for each city. Using the city shape information, we query local Open Street Map (OSM) elements matching specific tag-value pairs, obtaining the total number of POI and the aggregated area (in km$^2$) and length (in km) of all elements.

Within the OSM elements we analyze two elements: "nodes" and "ways". Within each OSM element there is a different set of categories, each of which can have different values (for instance, a public school is tagged in the "amenity" category with the value "school"). We selected 18 categories within "nodes" and "ways" that depict geographical, demographic and socio-economic features of the different municipalities.  

The table below lists the specific queries made for each category.

\begin{center}
\begin{longtable}{ |c|c|c|c|c| } 
\hline
Category & Measure & Key & Value \\
\hline
\hline
\multirow{3}{3cm}{Food\_point (poly)} & \multirow{3}{2cm}{Count (Area)} & \multirow{2}{*}{amenity} & \multirow{2}{5cm}{restaurant, bar, cafe, fast\_food, pub, food\_court, ice\_cream, marketplace, biergarten} \\ 
&&& \\
&&& \\
\hline
\hline
\multirow{5}{3cm}{Health\_point (poly)} & \multirow{5}{2cm}{Count (Area)} & \multirow{2}{*}{amenity} & \multirow{2}{5cm}{hospital, clinic, dentist, doctors, nursing\_home, pharmacy, veterinary} \\ 
&&& \\
&&& \\ \cline{3-4}
& & healthcare & *all* \\ \cline{3-4}
& & building & hospital \\ \cline{3-4}
& & shop & herbalist, nutrition\_supplements \\
\hline
\hline
\multirow{1}{3cm}{Industrial\_landuse} & \multirow{1}{2cm}{Count} & landuse & industrial \\ 
\hline
\hline
\multirow{1}{3cm}{Main\_road\_line} & \multirow{1}{2cm}{Length} & highway & *all* \\ 
\hline
\hline
\multirow{4}{3cm}{Natural\_landuse} & \multirow{4}{2cm}{Area} & landuse & village\_green, greenfield \\  \cline{3-4}
& & leisure & garden, park, dog\_park \\ \cline{3-4}
& & amenity & grave\_yard \\ \cline{3-4}
& & natural & wood, scrub, heath, grassland \\
\hline
\hline
\multirow{1}{3cm}{residential\_landuse} & \multirow{1}{2cm}{Count} & lanudse & residential \\ 
\hline
\hline
\multirow{1}{3cm}{commercial\_landuse} & \multirow{1}{2cm}{Count} & lanudse & commercial \\ 
\hline
\hline
\multirow{1}{3cm}{retail\_landuse} & \multirow{1}{2cm}{Count} & lanudse & retail \\ 
\hline
\hline
\multirow{3}{3cm}{retail\_point (poly)} & \multirow{3}{2cm}{Count (Area)} & lanudse & retail \\  \cline{3-4}
&& building & commercial, kiosk, retail, supermarket \\ \cline{3-4}
&& shop & *all* \\
\hline
\hline
\multirow{6}{3cm}{school\_point (poly)} & \multirow{6}{2cm}{Count (Area)} & \multirow{3}{*}{amenity} & \multirow{2}{5cm}{school, college, university, driving\_school, kindergarten, university, library, language\_school, music\_school} \\ 
&&& \\
&&& \\
&&& \\ \cline{3-4}
&& building & college, school, university, kindergarten \\ \cline{3-4}
&& landuse & education \\
\hline
\hline
\multirow{8}{3cm}{transport\_point (poly)} & \multirow{8}{2cm}{Count (Area)} & \multirow{3}{*}{amenity} & \multirow{2}{5cm}{bicycle\_parking, bicycle\_rental, boat\_rental, bus\_station, car\_rental, ferry\_terminal, motorcycle\_parking, parking, taxi} \\ 
&&& \\
&&& \\
&&& \\
&&& \\ \cline{3-4}
&& aeroway & *all* \\ \cline{3-4}
&& public\_transport & *all* \\ \cline{3-4}
&& building & train\_station \\
\hline
\hline
\multirow{11}{3cm}{entertainment\_point (poly)} & \multirow{11}{2cm}{Count (Area)} & \multirow{4}{*}{amenity} & \multirow{4}{5cm}{arts\_centre, casino, cinema, community\_centre, conference\_centre, events\_venue, gambling, nightclub, planetarium, social\_centre, theatre} \\ 
&&& \\
&&& \\
&&& \\
&&& \\
&&& \\ \cline{3-4}
&& historic & *all* \\ \cline{3-4}
&& leisure & *all* \\ \cline{3-4}
&& sport & *all* \\ \cline{3-4}
&& tourism & *all* \\ \cline{3-4}
&& building & civic \\
\hline
\caption{\textbf{Geographical categories and query details}. For each geographical category, we detail the measure (number of points, total area in km$^2$ or length in km) and the OSM key-value used in the query.}
\label{t.poi}
\end{longtable}
\end{center}

\clearpage 
\subsection{Prediction performance}

\begin{table}[h!]
    \centering
   
\begin{tabular}{llllllll}
\toprule
Metric & Model  & New York & Massachusetts & California &  Florida & Washington &   Texas \\
\midrule
CPC & Gravity pow &    0.545 &         0.412 &      0.538 &    0.548 &       0.43 &   0.469 \\
         & Gravity exp &    0.403 &         0.439 &      0.367 &    0.403 &      0.345 &   0.297 \\
         & Radiation &     0.43 &         0.556 &      0.492 &    0.482 &      0.455 &   0.471 \\
         & BMS Plausible &    0.551 &         0.544 &      0.026 &    0.533 &      0.588 &   0.524 \\
         & BMS Ensemble &    0.548 &         0.447 &        0.0 &    0.506 &      0.573 &   0.489 \\
         & BMS Predictive &    0.548 &         0.448 &        0.0 &    0.506 &      0.574 &   0.489 \\
         & Random Forest &    0.547 &         0.573 &      0.401 &    0.489 &      0.532 &   0.486 \\
         & Deep Gravity &    0.528 &         0.495 &      0.567 &    0.624 &      0.548 &    0.45 \\\hline
AbsErr & Gravity pow &  138.135 &        405.42 &    220.748 &  141.447 &    423.065 &  83.991 \\
         & Gravity exp &   150.11 &       462.597 &    298.928 &  178.304 &    459.965 &  91.784 \\
         & Radiation &  117.952 &       158.143 &    163.691 &    94.98 &    181.743 &  79.395 \\
         & BMS Plausible &   87.345 &       129.145 &     94.944 &   64.162 &    150.607 &   47.28 \\
         & BMS Ensemble &   92.037 &       129.399 &    100.663 &   67.566 &    153.518 &  48.914 \\
         & BMS Predictive &   91.926 &       129.899 &    100.922 &   67.694 &    153.703 &  48.915 \\
         & Random Forest &   92.059 &       105.783 &    102.838 &   68.671 &    123.555 &  52.184 \\
         & Deep Gravity &   89.483 &       129.293 &    141.101 &   75.972 &    143.069 &   75.79 \\\hline
RE & Gravity pow &    0.776 &         0.985 &      0.835 &    0.858 &      1.126 &   0.866 \\
         & Gravity exp &    0.812 &         0.994 &      0.935 &    0.926 &      1.088 &   0.882 \\
         & Radiation &    0.947 &          0.86 &      0.948 &     0.94 &      0.921 &   0.976 \\
         & BMS Plausible &    0.605 &         0.546 &      0.541 &    0.531 &      0.598 &   0.553 \\
         & BMS Ensemble &    0.612 &         0.568 &      0.535 &     0.54 &      0.605 &   0.534 \\
         & BMS Predictive &    0.612 &          0.57 &      0.535 &     0.54 &      0.605 &   0.534 \\
         & Random Forest &    0.597 &         0.504 &      0.566 &    0.552 &      0.569 &   0.547 \\
         & Deep Gravity &    0.692 &         0.697 &      0.793 &     0.72 &       0.72 &   0.962 \\\hline
LogRatio & Gravity pow &    0.942 &         1.234 &      0.994 &     1.12 &      1.162 &   1.289 \\
         & Gravity exp &    1.004 &         1.314 &      1.239 &    1.321 &      1.309 &   1.373 \\
         & Radiation &    2.575 &         1.598 &      2.552 &    2.397 &      2.078 &    3.04 \\
         & BMS Plausible &    0.607 &          0.56 &       0.55 &    0.551 &       0.62 &   0.552 \\
         & BMS Ensemble &     0.62 &          0.57 &      0.555 &    0.568 &      0.618 &   0.546 \\
         & BMS Predictive &    0.618 &         0.569 &      0.555 &    0.567 &       0.62 &   0.545 \\
         & Random Forest &    0.608 &         0.512 &      0.588 &    0.568 &       0.59 &    0.55 \\
         & Deep Gravity &    1.006 &         1.093 &      1.413 &    1.113 &      0.991 &   2.888 \\
\bottomrule
\end{tabular}
 \caption{\textbf{Summary of median values in Fig. 3 of the main text} Median of the prediction error for each metric, model and state.}
\label{table_fig3}
\end{table}

\clearpage

\section{Supplementary Figures}

\begin{figure*}[!htbp]
\centering
\includegraphics[width=0.8\textwidth]{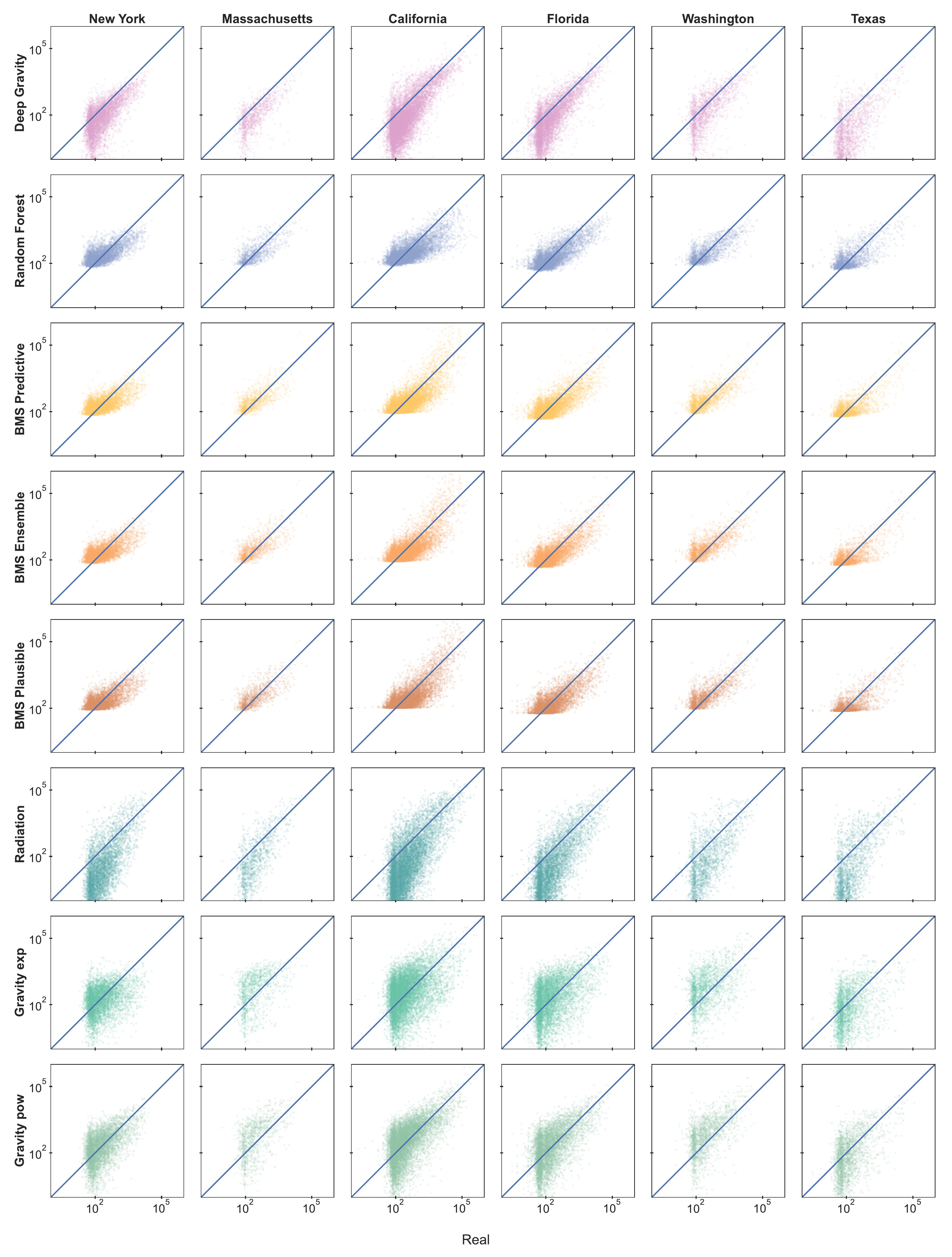}
\caption{\textbf{Model predictions of flows between municipalities. } Each panel shows the scatter plot of predicted flows between locations versus real flow values  (in logarithmic scale) for different states in the US (columns). Plots show results for the full set of models we consider (see Methods for a complete description of the models). Note that, we limit the range of predicted values to the range $[10^0, 10^6]$, but some models have predictions of flows $<1$ (Fig.~S2). }
\label{S1}
\end{figure*}

\begin{figure*}[!ht]
\centering
\includegraphics[width=0.95\textwidth]{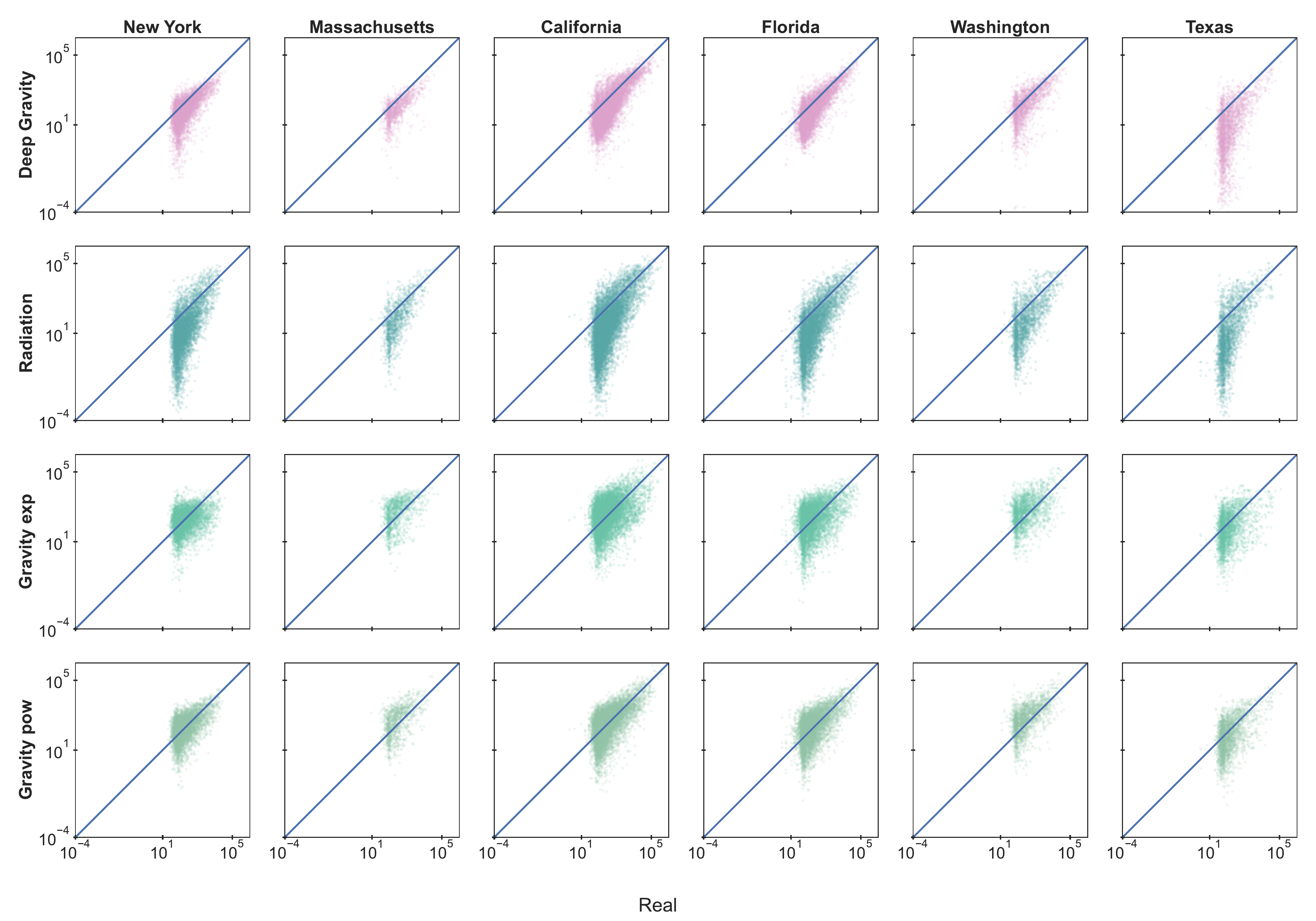}
\caption{\textbf{Model predictions of flows between municipalities. } For the models in Fig.~2, each panel shows the scatter plot of predicted flows between municipalities versus real flow values  (in logarithmic scale) for different states in the US (columns). Plots show the full range of predicted values (see Methods for a complete description of the models). Note how these three models underestimate by several orders of magnitude flows $\in[1,10]$.}
\label{S2}
\end{figure*}

\begin{figure*}[b!]
\centering
\includegraphics[width=0.95\textwidth]{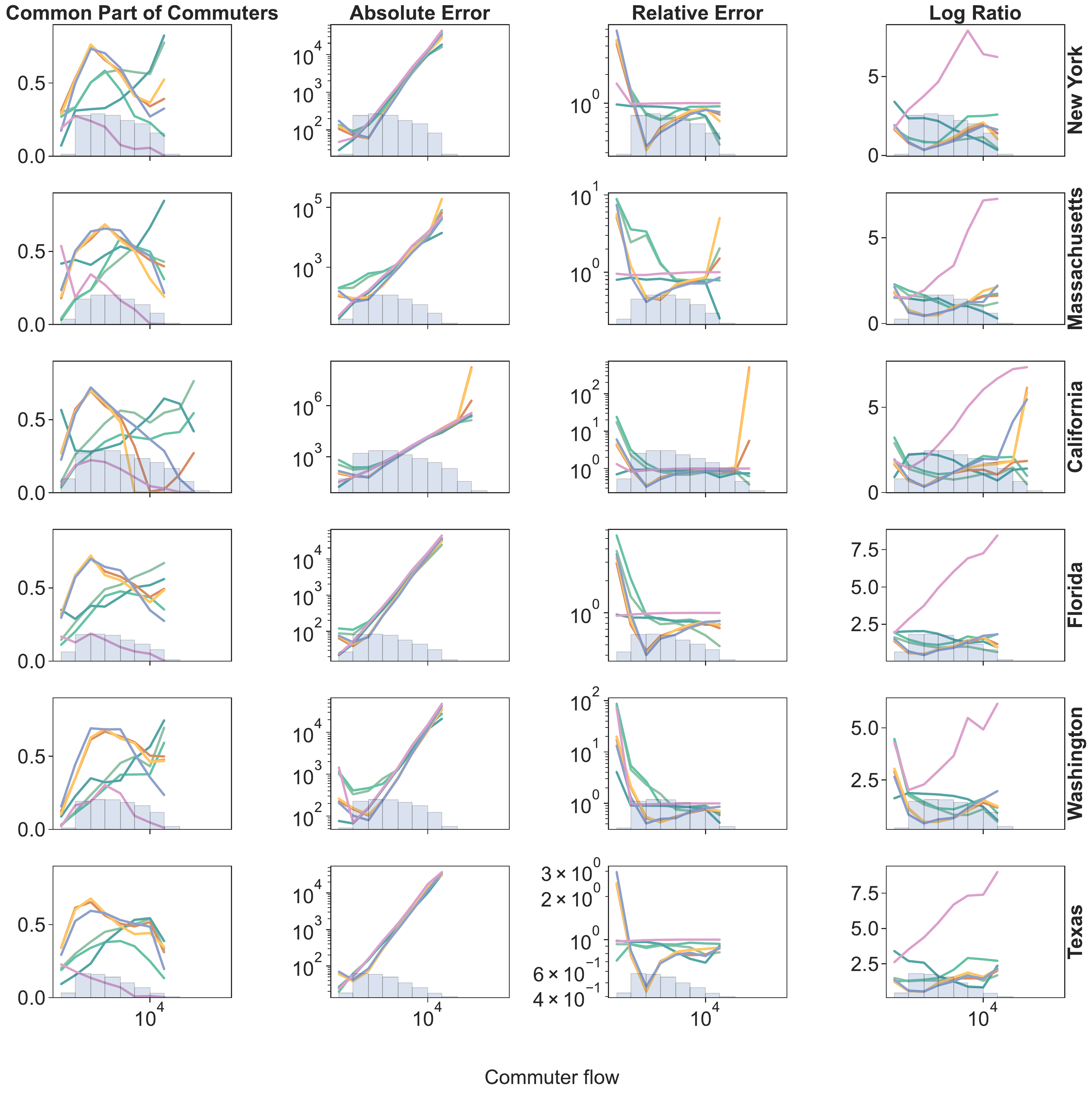}
\caption{\textbf{Performance for different flow ranges.} The observed flows between municipalities in the test set span six orders of magnitude and are  distributed in each state as shown by the histograms. For prediction of test set flow in each bin in the histogram, we show the predictive performance of different modeling approaches (lines, colors are the same as in Fig. S2) according to different metrics (columns). }
\label{S3}
\end{figure*}

\begin{figure*}[h!]
\centering
\includegraphics[width=0.95\textwidth]{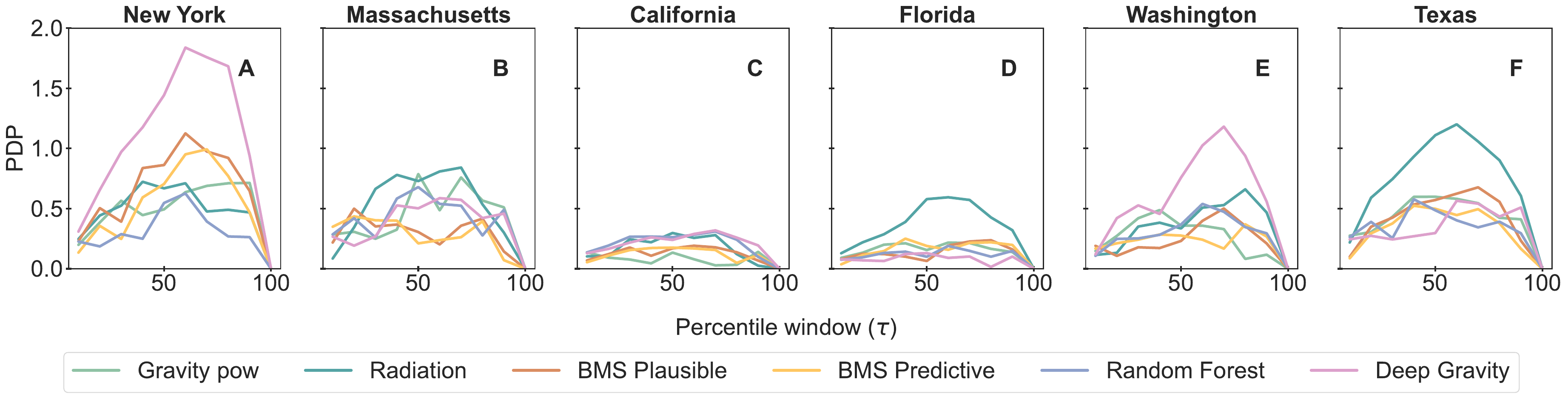} 
\caption[B]{\textbf{Model fairness across economic groups.} For each state, we classify municipalities as rich or poor based on the median income per capita (see Data - Economic Classification). We show the Partial Demographic Parity (PDP) for each country and model as a function of the percentile window (Methods).}
\label{S4}
\end{figure*}

\begin{figure*}[h!]
\centering
\includegraphics[width=0.95\textwidth]{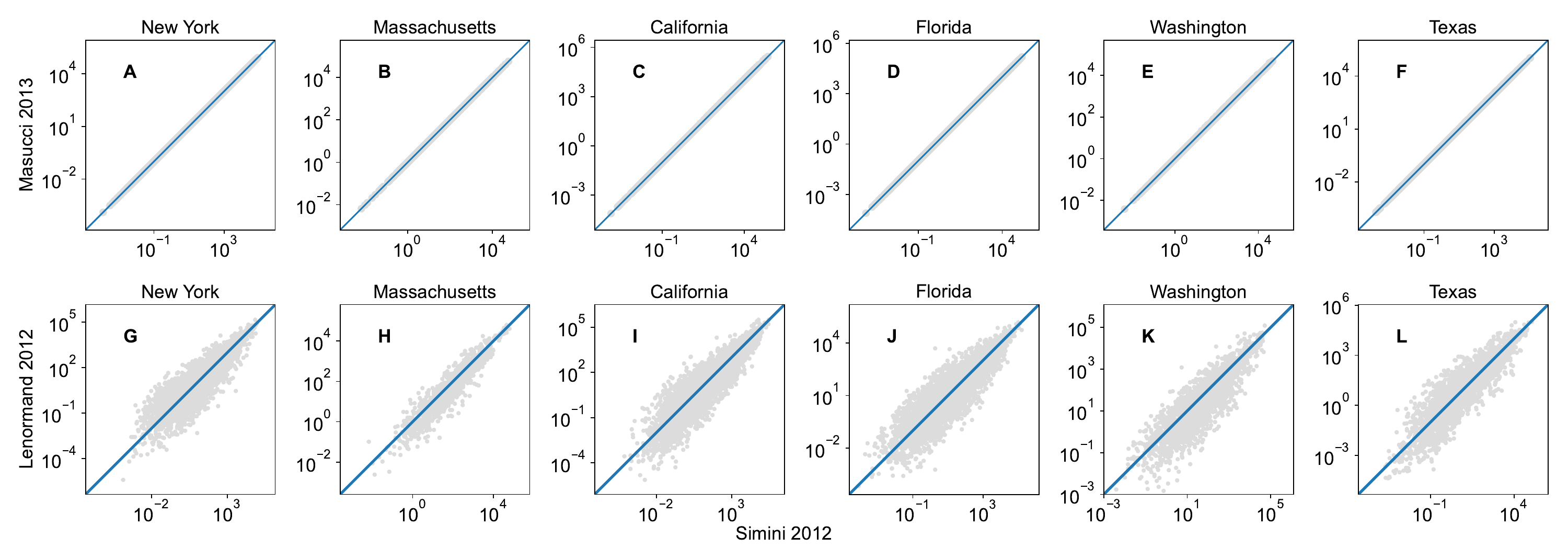} 
\caption[B]{\textbf{Variations of the radiation model.} For each state, we plot the predictions of a variation of the radiation model versus the predictions of the original radiation model. \textbf{A-F} show corrections introduced in Masucci et al. \cite{masucci2013}, where the flow of the original radiation is divided by $1-\frac{m_i}{\sum_i m_i}$. \textbf{K-L} show corrections introduced in Lenormand et al. \cite{lenormand2012}, where commuter flow is used instead of the municipality population size. As we can see four our dataset, these variations result in very similar predicitons of flows between municipalities, and, because of this we consider the original version of the model in our analysis thorughout the manuscript.}
\label{S5}
\end{figure*}

\clearpage

\bibliographystyle{naturemag}
\bibliography{mobility}